\documentclass[prb,nofootinbib,twocolumn]{revtex4} 

\usepackage{graphicx}
\usepackage{dcolumn}
\usepackage{bm}
\usepackage{threeparttable}
\usepackage{times}
\usepackage{mathptmx}
\usepackage{lscape}
\usepackage{natbib}
\usepackage{amsmath}
\usepackage{amssymb}
\usepackage{braket}
\usepackage{comment}
\usepackage{color}

\def\degree{\kern-.2em\r{}\kern-.3em}

\begin{document}

\title{ A Transport-Information Framework for Continuous-Fiber-Aligned Coarse-Graining Representation Sets  }

\author{Koretaka Yuge$^{\dagger}$}
\affiliation{
Department of Materials Science and Engineering,  Kyoto University, Sakyo, Kyoto 606-8501, Japan\\
}%

\begin{abstract}
{ 
Consider representation sets of interest whose elements are associated with continuous realizations related to them through coarse-graining (CG). CG can be regarded as an operation that reduces a representation by discarding information that is not retained in its output. However, a systematic description of the information structure associated with CG in general remains to be established. Here, we propose a transport-information framework for continuous-fiber-aligned representation sets, in which each representation is associated with a continuous realization through a fiber structure, with CG defining the projection onto the representation. Within this framework, CG is described through a local-map operation that transports continuous probability mass onto representative points. The resulting transport cost is expressed as the contraction of the second-moment structure of the underlying continuous representation with the transport metric. Building on the previously proposed framework of unavoidable canonical nonlinearity (UCN), this construction provides a non-arbitrary information-theoretic interpretation of the CG-associated transport cost in terms of information indistinguishability, quantified by an expected second-order structure on statistical manifold associated with the CG geometry. We also show that the proposed framework can also be naturally extended to the CG of the discrete measures themselves. The present framework provides a basis for describing information costs associated with individual CG operations and for comparing representation sets while accounting for their underlying continuous realizations.
}
\end{abstract}

\maketitle

\section{Introduction}
Representations provide a means of describing systems in terms of the information relevant to a given purpose. Depending on the representation, different aspects of a system may be retained, while others are reduced or discarded. Consider a representation  obtained from a coarse-graining (CG) operation on underlying continuous description: The resulting representation does not generally preserve all the information about the original one. Such a relationship between representations and their continuous realizations raises a fundamental question: how can the \textit{intrinsic} information structure associated with the generation of a representation through CG be characterized?
The important point is that, again,  a representation obtained through CG does not, in general, fully specify the information about the CG operation. In particular, different set of continuous realizations and CG operations may lead to the same, or equivalent, representations.  Consequently, an analysis based solely on the resulting representations may fail to capture the operational structure associated with their generation. This issue is not limited to comparisons between different representations. Even for a single given representation, the CG operation could induce characteristic transition cost and an associated information structure. A general framework should therefore account for both the costs associated with individual representation-generating operations and the comparison of representations in terms of their underlying continuous realizations and CG operations.

Generally, such CG operation conceptually acts on the sample space, which cannot be directly accessed by information-theoretic way unless introducing arbitrary structures. 
A useful starting point to overcome this problem is employing the non-arbitrary relationship between transport cost and information-geometric indistinguishability due to the discretization, established in the framework of unavoidable canonical nonlinearity (UCN)\cite{UCN} developed by our previous work. In the UCN, a local-map (LM) scheme transports probability mass from a continuous distribution to representative discrete points as the center of any convex bounded set. In the setting considered in UCN, this transport structure is related to the expected Kullback-Leibler (KL) divergence\cite{KL} between nearby distributions, providing an information-theoretic interpretation of the transport cost in terms of statistical indistinguishability. 
Importantly, the relevant geometric structure arises from the LM operation itself; the discretization-cell geometry in the original formulation provides one way of specifying the corresponding transport costs.

The representation-aware transport-information (DRA) measure\cite{DRA} in our subsequent work extends this perspective to the comparison of discrete distributions whose underlying continuous distributions and discretization schemes need not coincide. Rather than comparing only the final discrete distributions, DRA evaluates the KL divergences between continuous distributions averaged over parameter variations induced by their discretization schemes, which is non-arbitrarily characterized through the transport-information correspondence in the UCN. This construction allows the comparison to retain information about the continuous distributions as well as the discretization operations underlying the representations. 

These developments motivate a broader framework in which representations (not always as discrete distributions) are treated as elements of representation sets, each conceptually associated with a continuous realization and its CG operation through a fiber structure: The continuous realizations form the total space, while the representation set serves as the base space. The projection from the total space to the base space is defined by the composition of the CG operation and a one-to-one mapping. This formulation separates the information-reducing operation from the choice of how its output is represented, while retaining the continuous realization associated with each representation.
In this work, we propose a transport-information framework for continuous-fiber-aligned representation sets, termed DRAC. The framework uses the the construction to characterize the transport cost associated with an individual CG operation and incorporates the information-geometric interpretation established by UCN. It also provides a basis for comparing representation sets while accounting for their underlying continuous realizations and information-theretic costs for CG. Thus, DRAC places individual CG costs and representation-comparison costs within a common conceptual framework. Its central perspective is that the transport-information structure of a representation is not determined solely by the resulting representation, but also by the continuous realization and the operation through which that representation is obtained. The details are discussed in the following sections.

\section{Concept and Construction}
\subsection{Representation Sets and Continuous-Fiber Structure}

Consider a set of representations of interest, denoted by $\mathcal{R}=\{R_k\}_{k\in K}$, where $K$ is an index set. The elements of $\mathcal{R}$ need not be probability distributions or discrete configurations. Rather, each $R_k$  represents an object in a form appropriate to the purpose of interest, and is assumed to be associated with an underlying continuous realization and its coarse-graining (CG) operation.
More concretely, let $\mathcal{E}$ denote the space of these continuous realizations. For each $R_k$, the corresponding fiber $\mathcal{F}_k\subseteq\mathcal{E}$ represents the corresponding continuous realization. The collection of fibers defines a total space $\mathcal{E}$ over the representation set $\mathcal{R}$. In this construction, the representation set of interest serves as the base space, while the continuous realizations constitute its fiber structure.

To formalize this correspondence, let $\mathcal{B}_{\mathrm{CG}}$ denote the set of points obtained by applying CG to the continuous realizations in $\mathcal{E}$. We here introduce a bijective map $\Gamma:\mathcal{B}_{\mathrm{CG}}\to\mathcal{R}$, which identifies each CG output with its corresponding representation of interest. The projection from the total space to the representation set is then defined by the composite map:
\begin{eqnarray}
\pi_{\mathcal{R}}=\Gamma\circ\pi_{\mathrm{CG}} \qquad \pi_{\mathcal{R}}:\mathcal{E}\to\mathcal{R},
\end{eqnarray}
where $\pi_{\mathrm{CG}}:\mathcal{E}\to\mathcal{B}_{\mathrm{CG}}$ denotes the CG operation. 
The information reduction considered in this framework is associated with the CG itself, whereas $\Gamma$ specifies how the resulting CG points are identified with the representations of interest. This distinction allows the base space to be chosen according to the representational context without requiring it to coincide with the set of CG output points.

The term ``continuous-fiber-aligned representation set'' refers to a representation set equipped with this correspondence between its elements and their underlying continuous realizations with CG. The continuous realization and its associated CG operation provide the structure from which transport-information costs can be defined, as developed in the following subsections.

\subsection{Coarse-Graining and Local-Map Transport Cost}

We now formulate the coarse-graining (CG) operation introduced in terms of local-map (LM) transport. Consider a continuous probability distribution $P_k$ associated with a representation $R_k$, defined on a sample space 
$\mathcal X_k$. Let $x_k^0\in\mathcal X_k$ be the representative point  (as the center of $P_{k}$) onto which the probability mass is aggregated by CG: In the present framework, the CG operation on a continuous representation is defined as aggregating all probability mass onto its center, thereby discarding information about the distribution beyond its center.
The LM operation $T$ is defined by the deterministic map: 
\begin{eqnarray}
T_k(x)=x_k^0,\qquad x\in\mathcal X_k,
\end{eqnarray}
which transports all probability mass of \(P_k\) to \(x_k^0\). The resulting coarse-grained distribution is the point mass of
$\delta_{x_k^0}$.
In the UCN framework, the transport cost is measured based on 2-Wasserstein distance,\cite{WS} where its metric $G_{k}$ of the cost function is specially selected discussed later. Under the above LM scheme, the corresponding cost can be explicitly given by
\begin{eqnarray}
C_{\mathrm{LM}}^{(k)} = \int_{\mathcal X_k} (x-x_k^0)^{\mathsf T}  G_k  (x-x_k^0)\,P_k\left(x\right) dx.
\end{eqnarray}
Defining the second-moment matrix relative as
\begin{eqnarray}
M_k =\int_{\mathcal X_k} (x-x_k^0)(x-x_k^0)^{\mathsf T}\,P_k\left(x\right)dx,
\end{eqnarray}
the transport cost for the CG can be expressed as the following contruction:
\begin{eqnarray}
C_{\mathrm{LM}}^{(k)} =\textrm{Tr}(G_kM_k). 
\end{eqnarray}
This identity holds whenever the second moment is finite.

\subsection{Information-Geometric Interpretation of CG}
The transport cost defined in the previous section acquires an information-theoretic interpretation when the transport metric of $G_{k}$ is aligned with the Fisher metric for $P_{k}$. This connection is originally established in the framework of unavoidable canonical nonlinearity (UCN), which relates the geometric structure of a local-map (LM) operation from continuous to discrete distributions to the statistical indistinguishability of nearby probability distributions, without introducing any arbitrary structure. Note that in the UCN, the discretization is originally performed on the infinite number of discrete supports associated with the translation of any bounded convex set, while the present CG explicitly introduce a \textit{single} discrete support: For both cases, under the LM-scheme transport plan, the resultant cost is eventually formulated by the same contruction form between the transport metric and the second-order moment structure $M$, where the original UCN adopts $M$ as the second-moment of the discretization cell while the present construction $M$ as the second-moment of the continuous distribution $P_{k}$. These conceptual differences in the orignal UCN and the present CG operation requires additional consideration about adopting the Fisher metric as transport cost, including the distinction between its structural role and the choice of an overall scale: They are discussed separately in Sec. *Canonical Choice of the Transport Metric*.

Based on the UCN framework, the admissible Fisher metric is not arbitrary, but the underlying parameter on statistical manifold should be identified with the support coordinates. In the present construction, we specifically choose such parameters $\xi$ as the translation family for $P_{k}$, namely, 
\begin{eqnarray}
P_{k:\xi}(x)=P_k(x-\xi).
\end{eqnarray}
Let \(\Omega_k\) denote the Fisher information matrix of this family, evaluated at the reference parameter. Under the usual regularity conditions, the Kullback--Leibler (KL) divergence between a reference distribution and a nearby translated distribution has the local expansion
\begin{eqnarray}
D_{\mathrm{KL}}
\bigl(P_{k:\xi+\delta\xi}:P_{k:\xi}\bigr)
=
\frac{1}{2}\delta\xi^{\mathsf T}
\Omega_k\delta\xi
+o(\|\delta\xi\|^2).
\end{eqnarray}
Thus, the Fisher information matrix determines the local quadratic form governing the statistical distinguishability of nearby members of the translation family. Define this quadratic form by
\begin{eqnarray}
Q_k(\delta\xi)
=
\frac{1}{2}\delta\xi^{\mathsf T}
\Omega_k\delta\xi.
\end{eqnarray}
In general, \(Q_k\) is the second-order approximation to the KL divergence, rather than the KL divergence itself at a finite parameter displacement.

Next, let \(\rho_k\) be a centered distribution of parameter displacements satisfying
\begin{eqnarray}
\mathbb E_{\rho_k}[\delta\xi]=0,
\qquad
\mathbb E_{\rho_k}
[\delta\xi\,\delta\xi^{\mathsf T}]
=M_k.
\end{eqnarray}
The expected local quadratic form is then
\begin{eqnarray}
\begin{aligned}
\mathbb E_{\rho_k}[Q_k(\delta\xi)]
&=
\frac{1}{2}
\mathbb E_{\rho_k}
\left[
\delta\xi^{\mathsf T}
\Omega_k\delta\xi
\right]\\
&=
\frac{1}{2}\operatorname{Tr}(M_k\Omega_k).
\end{aligned}
\end{eqnarray}
Combining this result with the transport-cost expression obtained above, with \(G_k=\Omega_k\), gives
\begin{eqnarray}
C_{\mathrm{LM}}^{(k)}
=
2\,\mathbb E_{\rho_k}[Q_k(\delta\xi)].
\end{eqnarray}
This identity is an exact correspondence between the quadratic transport cost and the expected local Fisher quadratic form, provided that the displacement covariance is \(M_k\). It does not require the CG operation itself to be realized as a physical random displacement process: \(\rho_k\) provides a representation of the second-moment structure of the aggregation operation in the tangent space of the statistical family.

The relation to the KL divergence must be interpreted locally. When the parameter displacements are confined to a sufficiently small neighborhood of the reference parameter, the expected KL divergence is approximated by the expected Fisher quadratic form. 
Accordingly, the correspondence established here concerns the common local quadratic structure; it does not, by itself, identify the full finite CG transport cost with twice the expected KL divergence for an arbitrary displacement distribution.

This construction is related to the UCN framework, in which the local information-geometric structure and the transport cost associated with continuous-to-discrete representation are connected through a discretization-scale limit. The geometric origins of the second-moment matrix differ between the two settings: in UCN, it is determined by the discretization cell, whereas in the present CG construction it is determined by the continuous distribution and the selected representative point. The present formulation therefore extends the transport-information correspondence to a single-point aggregation operation without requiring the discretization-cell construction used in UCN. The rationale for using the Fisher metric on the transport side is considered next.

\subsection{Canonical Choice of the Transport Metric}
The preceding construction reveals a dual aspect of the CG operation as transport cost on the sample space and the expected squared norm of the tangent vector associated with the continuous representation on statistical manifold. On the one hand, CG is defined as a sample-space operation, and its operational cost is directly governed by the transport metric \(G_k\). In this sense, the transport side is primary. On the other hand, the same operation is associated, at least on the information-geometric side, with the local statistical indistinguishability of the translation family around the reference distribution \(P_k\), as characterized by its Fisher information matrix \(\Omega_k\). The choice of transport metric should account for both aspects without confusing their respective roles.

Because the transport cost is primary to the operational definition of CG, the transport metric should retain the information-geometric second-order structure associated with the operation, rather than suppressing directions in which the translation family has nonzero Fisher sensitivity. A necessary condition for this requirement is
\begin{eqnarray}
\ker G_k\subseteq\ker\Omega_k.
\end{eqnarray}
Indeed, any direction that is null with respect to the transport metric must also be null with respect to the Fisher information matrix; otherwise, the transport metric would discard a direction that remains distinguishable in the information-geometric description. This condition implies the rank inequality
\begin{eqnarray}
\operatorname{rank}(G_k)
\geq
\operatorname{rank}(\Omega_k).
\end{eqnarray}
The kernel inclusion expresses the requirement that the transport metric should not lose Fisher-sensitive directions. It is a necessary condition for retaining the information structure, rather than a sufficient condition that uniquely determines the transport metric.

In particular, if the Fisher information matrix is full rank, then
\begin{eqnarray}
\ker\Omega_k=\{0\}.
\end{eqnarray}
The above condition consequently requires
\begin{eqnarray}
\ker G_k=\{0\},
\qquad
\operatorname{rank}(G_k)=\dim\mathcal X_k.
\end{eqnarray}
Thus, when the translation family is locally distinguishable in every parameter direction, a transport metric with degenerate directions cannot retain its full Fisher information structure.

These constraints do not, by themselves, require \(G_k=\Omega_k\). For example, a positive scalar multiple of the Fisher information matrix satisfies the same kernel and rank conditions. Such a scalar changes the overall scale of the transport cost while preserving the relative weighting of directions. It can therefore be interpreted as a conversion factor between the units used on the transport and information-geometric sides. In the absence of an independently prescribed transport-cost unit, this overall scale may be fixed to unity as a normalization convention.

Beyond this scalar freedom, more general transport metrics may introduce additional directional weights or transformations. Such metrics are mathematically admissible, but the additional structure is not specified by the Fisher information matrix alone. Adopting it would therefore require an independent justification. To avoid such additional structure, we select the transport metric by directly adopting the Fisher quadratic structure on the same displacement space, without transforming or reweighting its directions:
\begin{eqnarray}
G_k=\Omega_k.
\end{eqnarray}
This is the canonical choice in the present framework. Its rationale is not that alternative metrics are mathematically forbidden, nor that the kernel and rank conditions uniquely determine the Fisher metric. Rather, the choice follows from the operational duality of CG, the requirement to retain the associated information-geometric structure, and the adoption of that structure without additional transformations, after fixing the overall scale convention.

The reasoning is consistent with the motivation underlying UCN, although the specific constructions differ. In UCN, the relation between transport and information geometry is established through the limiting structure associated with discretization. Here, the relation is formulated directly for the single-point CG operation through its second-moment matrix and the Fisher quadratic form of the associated translation family. In both settings, the Fisher metric provides a canonical means of relating the operational structure on the sample space to local statistical distinguishability, without claiming absolute mathematical uniqueness among all possible transport metrics.

\subsection{Costs for Individual Representations and Representation Sets}

The preceding subsections establish the transport-information structure associated with an individual coarse-graining (CG) operation. We now place this structure within a framework that addresses both individual representations and comparisons between representation sets.

\subsubsection{Information cost associated with an individual representation.} 
This case is the most straightforward one. Let $\mathcal R=\{R_m\}_{m\in I}$ be a representation set equipped with the continuous-fiber structure introduced in Sec. II.A. For each $R_m$, let $P_{c,m}$ denote its associated continuous realization, and let $M_m$ and $\Omega_m$ denote the second-moment matrix associated with its CG operation and the Fisher metric of the corresponding statistical family, respectively. Under the conditions specified in Sec. II.C, the transport-information cost associated with the individual CG operation is defined as 
\begin{eqnarray}
C_m=\operatorname{Tr}(M_m\Omega_m). 
\end{eqnarray}
In the corresponding local information-geometric construction, this cost is related to the expected KL divergence for this family. Thus, each representation can be directly associated with a transport-information cost determined by its underlying continuous realization and the CG operation through which it is obtained.

\subsubsection{Pairwise comparison of representations} 
Consider two representation sets, $\mathcal R=\{R_m\}_{m\in I}$ and $\mathcal Q=\{Q_j\}_{j\in J}$, whose elements are associated with their continuous realizations $P_{c,m}$ and $Q_{c,j}$ and CG operations. To compare them, we adopt the representation-aware transport-information (DRA) measure previously developed. For a pair $(R_m,Q_j)$, its comparison can be defined by an average of KL divergences over the parameter-variations associated with the two continuous-reference representations:
\begin{widetext}
\begin{eqnarray}
c_{mj} = D_{\mathrm{RA}}(R_m:Q_j)=\iint \rho_m(\delta\xi)\,\nu_j(\delta\eta) D\bigl(P_{c,m:\xi+\delta\xi}:Q_{c,j:\eta+\delta\eta}\bigr) d\left(\delta\xi\right) d\left(\delta\eta\right). 
\end{eqnarray}
\end{widetext}
The comparison for a pair should admit a mathematically well-defined divergence-based comparison, as well as the physically (or conceptually) meaningful comparison. 
Note that when the continuous reference distributions do not satisfy the absolute-continuity conditions required for a finite KL divergence, a comparison based on KL may be infinite or otherwise unsuitable. In such cases, another well-defined measure, such as the Jensen-Shannon divergence,\cite{JS} would be adopted.
The $\rho$ and $\nu$ are, in a similar fashion to the previous discussions, chosen so that their covariance is aligned with the second-moment of the continuous distributions. In the UCN framework, the DRA has more explicit expression as the sum of divergence $D$ between $P_{c}$ and $Q_{c}$ and the trace contruction between second-moment of the continuous distributions and the Hessian of $D$: The former explicitly represents the information-theoretic difference in the underlying continuous distributions, while the  latter purely comes from the information-deformation due to the CG operation. 
Therefore, the pairwise cost retains information about the continuous realizations and their associated representation-generating operations that may not be recoverable from the final representations alone. 

\subsubsection{Selection and aggregation of comparisons} 
In general, not every pair of representations needs to be compared. Let
\begin{eqnarray}
\mathcal J\subseteq\mathcal R\times\mathcal Q
\end{eqnarray}
denote the set of pairs selected for comparison. The selection may be determined by the purpose of the analysis and by the mathematical admissibility of the corresponding continuous divergence-based comparisons. Assign a nonnegative weight $w_{mj}$ to each selected pair. When the total weight is positive, an aggregate comparison cost can be naturally defined as
\begin{eqnarray}
\langle C_{\mathrm{RA}}\rangle_{\mathcal J} = \displaystyle\sum_{(m,j)\in\mathcal J} w_{mj}c_{mj} \cdot 
\left(\displaystyle\sum_{(m,j)\in\mathcal J}w_{mj}\right)^{-1}
.
\end{eqnarray}
The weights can, for example, represent the relative importance of selected comparisons or the probabilities with which the pairs are realized.  
The selection of $\mathcal J$ and the weighting scheme are part of the specification of the comparison problem, rather than intrinsic properties of either representation set. This flexibility allows the framework to accommodate different comparison objectives without requiring a single universal rule for pairing representations.

\subsubsection{Scope of the framework} 
The resulting construction distinguishes three related but conceptually different objects: the transport-information cost associated with an individual CG operation, the pairwise comparison cost between representations, and an aggregate cost over a selected set of comparisons. These quantities share a common basis in the continuous realizations and their associated transport-information structures, but they are not interchangeable. In particular, the individual CG costs do not uniquely determine the pairwise comparison measure.

Within DRAC, the DRA construction supplies the pairwise comparison component, while the LM and UCN constructions provide the basis for characterizing the transport-information cost associated with individual CG operations. Their integration into a framework of continuous-fiber-aligned representation sets makes it possible to describe both the information structure accompanying representation generation and the information costs associated with comparing the resulting representations.

\subsection{Pushforward Comparison through CG-Induced Cost Distributions}
The direct comparison introduced above evaluates information-theoretic divergences between the continuous realizations associated with the CG operation, including their specific families of submanifolds. This construction retains information about the underlying continuous distributions, but it requires the relevant continuous distributions to admit a well-defined divergence-based comparison. When the associated sample spaces for the fiber pair have different dimensions or lack a natural correspondence between their coordinates, their direct comparison may become difficult to formulate without introducing additional arbitrary structure. We therefore introduce an alternative, unified comparison based on probability measures induced by the coarse-graining (CG) operation itself.

For each representation $R_m$, let $P_{c,m}$ be its associated continuous realization on a sample space $\mathcal X_m$, and let $x_m^0\in\mathcal X_m$ be the representative point to which probability mass is aggregated by CG. We define the pointwise LM transport cost by
\begin{eqnarray}
c_m(x)=(x-x_m^0)^{\mathsf T}\Omega_m(x-x_m^0),
\end{eqnarray}
where $\Omega_m$ is the prescribed Fisher metric. Under this definition, we consider the corresponding pushforward probability measure:
\begin{eqnarray}
\label{eq:pm}
\mu_m=(c_m)_\#P_{c,m}.
\end{eqnarray}
We can clearly see that its expectation gives the LM transport cost,
\begin{eqnarray}
\int_{\mathbb R_{\geq 0}}s\,\mu_m\left(s\right) ds = \int_{\mathcal X_m}c_m(x)\,P_{c,m}\left(x\right) dx = C_{\mathrm{LM}}^{(m)}.
\end{eqnarray}
Thus, the introduced pushforward measure retains not only the mean transport cost but also the local contribution to the transport cost or the local contribution to the statistical indistinguishability induced by the CG operation. 

Under these preparations, for two representations $R_m$ and $Q_j$, let $\mu_m$ and $\nu_j$ denote the respective CG-induced measures defined as Eq.~\eqref{eq:pm}. We define the pushforward comparison by
\begin{eqnarray}
D_{\mathrm{CG}}^{\mathrm{push}}(R_m:Q_j) = D(\mu_m:\nu_j),
\end{eqnarray}
where $D$ can be chosen as KL-divergence, JS-diverence or other information-theoretic divergences depending on the purpose of the setup. 
Here, the divergence is evaluated between the pushforward measures on the cost axis, rather than between the original continuous realizations or the final representations. The comparison therefore characterizes differences between the operational cost distributions induced by the two CG procedures. Since the comparison takes place on the common cost axis, it can be formulated even when the original continuous sample spaces have different dimensions or lack a natural coordinate-wise correspondence, provided that the cost variables are defined on a common scale.

This construction is distinct from the direct comparison of continuous distributions described above. The direct construction compares the underlying statistical realizations, with their associated parameter-variation families, whereas the pushforward construction compares the distributions of an operation-induced quantity. Consequently, the latter need not retain all information about the original continuous distributions: distinct realizations may induce the same cost distribution. This loss of information reflects the choice to compare representations through the operational structure captured by the specified CG cost.

The construction can also be applied to discrete measures whose final supports are not mutually inclusive. Let $P_A$ and $P_B$ be discrete distributions supported on $\{A_i\}_{i\in I}$ and $\{B_k\}_{k\in K}$, respectively. Their continuous realization and CG-induced cost distribution are characterized by the pushforward measures of $\mu_{A_i}$ and 
$\mu_{B_k}$.
Given a normalized nonnegative pair-weighting distribution $\Pi_{ik}$, the aggregate comparison between $P_{A}$ and $P_{B}$ can be given by
\begin{eqnarray}
D_{\mathrm{CG}}^{\mathrm{push}}(P_A:P_B) = \sum_{i\in I}\sum_{k\in K} \Pi_{ik} D(\mu_{A_i}:\mu_{B_k})
\end{eqnarray}
For independently sampled support states, one possible choice is $\Pi_{ik}=P_A(A_i)P_B(B_k)$; other pair-weighting schemes may be adopted when the comparison objective specifies a different pairing structure. If no probability-based weighting is available or intended, one may instead sum the pairwise costs over a selected set of pairs without interpreting the result as a normalized average.

The pushforward construction thus provides an alternative route for comparing continuous-fiber-aligned representation sets. It employs the CG operation to induce a common observable space for comparison. Both constructions fit within the proposed transport-information framework, but they retain different aspects of the information associated with representation generation.

\subsection{Coarse-Graining of Continuous Geometric Objects}
The coarse-graining (CG) framework described above can be naturally extended from probability distributions to the discretization of continuous geometric objects. Consider a filled geometric object $G\subseteq\mathbb{R}^{d}$, and let 
$\{S_i\}$ be a collection of bounded regions that partition the ambient space, up to their boundaries. The portion of $G$ contained in the $i$-th region is defined as
\begin{eqnarray}
A_i=G\cap S_i.
\end{eqnarray}
The regions $S_i$ need not have identical shapes or sizes. For each $A_i$ with nonzero measure, we introduce a normalized continuous probability measure $P_i$ describing its local geometric structure. For example, with respect to the $d$-dimensional Lebesgue measure, a weighted measure can be defined as
\begin{eqnarray}
P_i(dx)=\frac{w_i(x)\mathbf{1}_{A_i}(x)}{\int_{A_i}w_i(z)\,dz}\,dx,
\end{eqnarray}
where $w_i(x)\geq 0$ is a weighting function and the normalization integral is assumed to be finite and strictly positive. This construction includes the uniform measure as a special case, while allowing the local geometric contribution to vary within each region. Importantly, the construction does not require assigning a single probability distribution to the entire object $G$: each portion $A_i$ is associated with its own normalized measure. When the Fisher information of the associated translation family is considered, the weighting function must also be chosen to satisfy the relevant regularity conditions. In particular, the behavior of the density near the boundary of $A_i$ can affect the finiteness of the Fisher information, as discussed below.
The representative point of each portion is defined by the mean of its associated measure,
\begin{eqnarray}
x_i^0=\int_{\mathcal X_i}x\,P_i(dx),
\end{eqnarray}
where $\mathcal X_i$ denotes the domain of the measure. Assuming that the first and second moments are finite, the CG operation maps $P_i$ to the point mass $\delta_{x_i^0}$. Repeating this operation over the portions with positive measure yields a discrete point representation $\{x_i^0\}$ of $G$. Portions with zero measure under the chosen weighting scheme do not define normalized measures and must therefore be excluded or treated separately. We refer to this procedure as the coarse-graining of a continuous geometric object into a discrete representation.

The local transport-information structure can be formulated in terms of the second-moment matrix about the representative point,
\begin{eqnarray}
M_i=\int_{\mathcal X_i}(x-x_i^0)(x-x_i^0)^{\mathsf T}\,P_i(dx).
\end{eqnarray}
Let $\Omega_i$ denote the Fisher information matrix of the associated translation family, provided that the relevant regularity conditions hold and that the Fisher information is adopted as the transport metric. The local LM transport cost is then expressed as
\begin{eqnarray}
C_i=\operatorname{Tr}(M_i\Omega_i).
\end{eqnarray}
This expression also admits an information-geometric interpretation in terms of the local quadratic structure of the statistical manifold. To make this interpretation explicit, let $\delta\xi$ denote a random parameter displacement whose second-moment matrix is $M_i$, so that
\begin{eqnarray}
M_i=\mathbb E_{\rho_i}[\delta\xi\,\delta\xi^{\mathsf T}].
\end{eqnarray}
Then the local cost can equivalently be written as
\begin{eqnarray}
C_i=\mathbb E_{\rho_i}\left[\delta\xi^{\mathsf T}\Omega_i\delta\xi\right].
\end{eqnarray}
Here, $\delta\xi$ induces a tangent vector to the statistical manifold of the translation family at the reference distribution, and $\delta\xi^{\mathsf T}\Omega_i\delta\xi$ is the squared norm of that tangent vector under the Fisher information metric. Thus, the local cost can be interpreted as the expected squared Fisher norm of the induced tangent vectors. This interpretation concerns the local quadratic structure defined by the Fisher information metric and does not, by itself, require the cost to admit a representation as the expectation of a divergence.

The use of the Fisher information matrix requires the associated translation family to satisfy appropriate regularity conditions, including the finiteness of the Fisher-information integral. For a measure supported on a bounded region, the behavior of its density near the boundary is particularly relevant. Choosing a density that vanishes sufficiently rapidly and regularly at the boundary can yield a finite Fisher information matrix, although the precise conditions depend on the density and domain. Alternatively, one may regularize a local measure, for example by convolution with a zero-mean Gaussian distribution, to obtain a smooth translation family with full support, subject to the relevant regularity conditions. Such regularization, however, extends the measure beyond $A_i$, and the resulting second-moment matrix includes the contribution of the smoothing distribution. The local measure should therefore be chosen consistently with the intended geometric representation and the associated transport-information cost. A detailed analysis of suitable measure constructions and their regularity conditions is our future scope.

Finally, the local costs can be combined to characterize the discretization of the entire geometric object. For example, an aggregate cost can be defined as
\begin{eqnarray}
C_{\mathrm{geom}}\left(G\right)=\sum_i\alpha_i C_i,
\end{eqnarray}
where $\alpha_i\geq 0$ specifies the relative contribution of each portion. The choice of $\alpha_i$ depends on the intended aggregate quantity: equal weights treat the local CG operations equally, whereas weights based on geometric measures or other prescribed criteria emphasize different aspects of the object. This extension illustrates that the DRAC framework can describe the transport-information structure of geometric discretization by associating each local portion with a continuous measure, a representative point, and a corresponding local CG cost.

\subsection{Coarse-Graining of a Fixed-Support Discrete Measures}
The preceding constructions consider a family of continuous distributions. We now describe an extended construction in which the reference distribution is itself discrete and the statistical family is defined on its fixed support. 

Consider a discrete probability distribution on an $f$-dimensional support space,
\begin{eqnarray}
P_d=\sum_{i=1}^{N}p_i^0\delta_{x_i},
\qquad
x_i\in\mathbb{R}^f,
\end{eqnarray}
where $p_i^0>0$ and $\sum_i p_i^0=1$. We assume that the minimum convex hull $\textrm{CV}$ of the supports is full-dimensional, i.e., 
\begin{eqnarray}
\dim\textrm{CV}\{x_1,\ldots,x_N\}=f.
\end{eqnarray}
This condition ensures that the support coordinates contain nondegenerate variation in every direction of the support space.
In a similar fashion to the continuous case, let the CG operation aggregate all probability mass onto the center:
\begin{eqnarray}
x_0=\sum_i p_i^0x_i.
\end{eqnarray}
For a constant positive-semidefinite transport metric $G$, the CG-induced transport cost is given by
\begin{eqnarray}
C_{\mathrm{LM}} =\sum_i p_i^0(x_i-x_0)^{\mathsf T} G(x_i-x_0) =\textrm{Tr}(MG),
\end{eqnarray}
where
\begin{eqnarray}
M=\sum_i p_i^0(x_i-x_0)(x_i-x_0)^{\mathsf T}
\end{eqnarray}
is the covariance matrix of $P_d$. This identity is exact and follows directly from the prescribed aggregation map. The metric $G$, however, is not uniquely determined by the transport operation alone.

To specify an information-geometric structure, we here introduce the following exponential family on the same fixed support:
\begin{eqnarray}
p_i(\xi)=
\frac{p_i^0\exp(\xi^{\mathsf T}x_i)}
{\sum_j p_j^0\exp(\xi^{\mathsf T}x_j)},
\qquad \xi\in\mathbb{R}^f.
\end{eqnarray}
The reference distribution $P_{d}$ is recovered at $\xi=0$, and the support points $x_i$ remain unchanged throughout the family. The pairing $\xi^{\mathsf T}x_i$ naturally identifies the parameter directions with the support-coordinate directions. Changes in $\xi$ reweight the fixed support according to the corresponding linear functions of $x$. Thus, the statistical family is aligned with the support axes without requiring the support itself to vary. In the present construction, continuity of the family and this coordinate alignment are imposed. Covariance under arbitrary elements of \(\mathrm{GL}(f)\), which would also transform the support, is not required as an additional condition, since this condition generally requires the changes of the discrete supports themselves.

Let $\Omega(\xi)$ denote the Fisher metric of this family. For the parameterization above, we obtain
\begin{eqnarray}
\Omega(\xi)=\operatorname{Cov}_{P_\xi}(X),
\end{eqnarray}
and hence, at the reference distribution,
\begin{eqnarray}
\Omega(0)=M.
\end{eqnarray}
The full-dimensionality assumption, together with the strictly positive weights, ensures that $\Omega(\xi)$ is positive definite for every finite $\xi$. We therefore adopt the Fisher metric at the reference point as the transport metric,
$G=\Omega(0)$.
This is an information-geometric choice within the specified class of support-aligned families, rather than a claim that the transport operation uniquely determines a Fisher metric independently of the family.

With this choice, the CG transport cost becomes
\begin{eqnarray}
C_{\mathrm{LM}}=\operatorname{Tr}(M\Omega(0)).
\end{eqnarray}
The same quadratic structure can be expressed on the statistical manifold. Let $\delta\xi$ be a centered stochastic  parameter displacement satisfying
\begin{eqnarray}
\mathbb{E}_{\rho}[\delta\xi]=0,
\qquad
\mathbb{E}_{\rho}[\delta\xi\,\delta\xi^{\mathsf T}]=M.
\end{eqnarray}
The expected squared Fisher norm of the corresponding tangent vectors is then
\begin{eqnarray}
\mathcal{I}_{\rho}
=\mathbb{E}_{\rho}\!\left[
\delta\xi^{\mathsf T}\Omega(0)\delta\xi
\right]
=\operatorname{Tr}(M\Omega(0))
=C_{\mathrm{LM}}.
\end{eqnarray}
Thus, under the stated covariance alignment, the LM transport cost is exactly equal to the expected local Fisher quadratic form. The displacement distribution $\rho$ specifies how the second-moment structure of the CG operation is represented in the tangent space; it need not describe a physical random displacement process. The present result therefore provides a fixed-support discrete realization of the transport-information correspondence, complementary to the continuous family construction discussed above.

\section{Conclusion}

In this work, we proposed a transport-information framework for continuous-fiber-aligned representation sets, in which representations are considered as their continuous realizations and the subsequent coarse-graining (CG) operations connecting them. Rather than treating a representation solely as a final output, the framework explicitly accounts for the information structure associated with its generation and for the continuous realizations.

The framework characterizes an individual CG operation through a local-map (LM) transport cost, which is expressed as the contraction of the second-moment structure of the underlying continuous distribution with a transport metric. When the transport metric is identified with the Fisher information metric and the corresponding statistical-family construction is adopted, this cost admits an information-theoretic interpretation in terms of expected KL divergence between continuous distributions, building on the previously established unavoidable canonical nonlinearity (UCN) framework. This provides a connection between the operational description of CG and the statistical indistinguishability for continuous probability distributions.

The framework also incorporates representation-aware transport-information (DRA) measures to compare representations through their underlying continuous realizations, rather than through their final representations alone. By distinguishing individual CG costs from pairwise comparison costs and their aggregation over selected pairs, the framework accommodates different objectives for analyzing representation generation and comparison within a common conceptual structure.

As a complementary construction, the framework can also be applied to discrete distributions on a fixed support whose convex hull is full-dimensional, which illustrates that the transport-information correspondence in CG operation can be formulated beyond continuous reference distributions.

The present framework establishes a basis for investigating information costs associated with CG and representation comparison beyond a single specific representation scheme. Further work will be needed to identify the conditions under which the construction can be extended to broader classes of CG operations and to develop concrete applications in physical and other representation systems.

\section{Acknowledgement}
This work was supported by Research Grant from Hitachi Metals$\cdot$Materials Science Foundation.


\begin{thebibliography}{9}
\bibitem{UCN} K. Yuge, J. Phys. Soc. Jpn. (in press).
\bibitem{KL} S. Kullback and R. A. Leibler, Ann. Math. Stat. 22, 79 (1951).
\bibitem{DRA} K. Yuge, arXiv: 2609.30944 [cond-mat.stat-mech].
\bibitem{WS} C. Villani, Optimal Transport: Old and New, Springer (2009).
\bibitem{JS} J. Lin, IEEE Transactions on Information Theory \textbf{37}, 145 (1991).
\end{thebibliography}
\end{document}